\documentclass[prl,twocolumn,floatfix,showpacs ]{revtex4-1}
\UseRawInputEncoding
\usepackage{amsmath}
\usepackage{amssymb}
\usepackage{graphicx}
\usepackage{dcolumn}
\usepackage{bm}
\usepackage[colorlinks=true,linkcolor=blue,anchorcolor=blue, citecolor=cyan,urlcolor=cyan]{hyperref}
\usepackage[mathlines]{lineno}
\usepackage{ulem}
\usepackage{epstopdf}
\begin{document}
\title{Electric-field-induced fully-compensated ferrimagnetism in experimentally synthesized monolayer MnSe}
\author{Liguo Zhang}
\affiliation{School of Electronic Engineering, Xi'an University of Posts and Telecommunications, Xi'an 710121, China}
\author{ Gangqiang Zhu}
\affiliation{School of Physics and Electronic Information, Shaanxi Normal University, Xi'an 716000, Shaanxi, China}
\author{San-Dong Guo}
\email{sandongyuwang@163.com}
\affiliation{School of Electronic Engineering, Xi'an University of Posts and Telecommunications, Xi'an 710121, China}
\begin{abstract}
Owing to their inherent characteristics of zero stray field and terahertz dynamics, two-dimensional (2D) zero-net-magnetization magnets demonstrate the potential for miniaturization, ultradensity and ultrafast performance.  Recently, fully-compensated ferrimagnet of  2D zero-net-magnetization magnets  has already attracted attention, as it can exhibit global spin-splitting, magneto-optical response  and  anomalous Hall effect [\textcolor[rgb]{0.00,0.00,1.00}{Phys. Rev. Lett. 134, 116703 (2025)}]. Therefore, it is very important to provide experimentally feasible strategies and materials to achieve fully-compensated ferrimagnets.
Here, we use the experimentally synthesized A-type $PT$-antiferromagnet (the joint symmetry ($PT$) of space inversion symmetry ($P$) and time-reversal symmetry ($T$)) MnSe as the parent material to induce fully-compensated ferrimagnetism through an out-of-plane electric field.   This electric field can remove the $P$ symmetry of the lattice, thereby breaking the $PT$ symmetry and inducing spin-splitting. When considering spin-orbital coupling (SOC), MnSe with an out-of-plane magnetization can achieve the anomalous valley Hall effect (AVHE). In addition, we also discuss inducing fully-compensated ferromagnetism via Se vacancies and Janus engineering. Our works can promote the further development of 2D fully-compensated ferrimagnets both theoretically and experimentally.
\end{abstract}
\maketitle
Magnetism is pivotal in driving technological advancements. Ferromagnets have been the focus of intensive research and have found widespread applications across various fields\cite{1a}. Compared with ferromagnets, zero-net-magnetization magnets offer more advantages for spintronic devices\cite{k1,k2}. They enable high storage density, robustness against external magnetic fields, and ultrafast writing speeds, all thanks to their zero-net magnetic moment.
From a symmetric perspective, the collinear magnets with zero-net-magnetization primarily consist of $PT$-antiferromagnets  (the joint symmetry ($PT$) of space inversion symmetry ($P$) and time-reversal symmetry ($T$)), alternamagnets, and fully-compensated ferrimagnets\cite{k4,k5}.

For $PT$-antiferromagnets, the energy bands are spin-degenerate throughout the entire Brillouin zone (BZ) ($E_{\uparrow}(k)$=$PT$$E_{\uparrow}(k)$= $E_{\downarrow}(k)$),  which  prevents many interesting physical effects, such as magneto-optical effect,  anomalous Hall effect and anomalous valley Hall effect (AVHE)\cite{zg1,zg2}.  For alternemagnets, the real-space magnetic order closely resembles that of a collinear antiferromagnet, characterized by a perfectly spin-compensated arrangement. However, in the momentum space, the spin-splitting represents a natural extension of ferromagnetism, featuring non-relativistic spin-split bands with $d$-wave, $g$-wave or $i$-wave symmetry\cite{k4,k5}.  Alternemagnets have exhibited a range of phenomena that are previously deemed exclusive to ferromagnets. These phenomena include non-relativistic lifted Kramers degeneracy,  anomalous Hall/Nernst effect, non-relativistic spin (polarized) currents, and magneto-optical Kerr effect\cite{zg1}.

Fully-compensated ferrimagnets represent a unique class of ferrimagnetic materials, characterized by their zero-net magnetization\cite{f1,f2,f3}. The spin-splitting in these materials occurs throughout the entire BZ  with $s$-wave symmetry, resembling the behavior observed in ferromagnets. Similar to alternemagnet, fully-compensated ferrimagnet can also show  anomalous Hall/Nernst effect, non-relativistic spin (polarized) currents, and  magneto-optical Kerr effect. Recently, the importance of two-dimensional (2D) fully-compensated ferrimagnetism has gained increasing recognition, thereby expanding the domain of low-dimensional spintronic materials\cite{f4}. Hence, it is crucial to develop experimentally viable strategies and materials for the realization of 2D fully-compensated ferrimagnets.   From the perspective of symmetry, fully-compensated ferrimagnet can be obtained by breaking the  inversion symmetry of $PT$-antiferromagnet and the rotational or mirror symmetry of alternemagnet.

A monolayer of MnSe, combined with an AA-stacked bilayer silicene crystal lattice and exhibiting A-type antiferromagnetic (AFM) ordering, has recently been successfully synthesized  experimentally\cite{zg3}.  MnSe is a  typical $PT$-antiferromagnet. Here, we use an out-of-plane electric field to break the $PT$ symmetry of MnSe, thereby achieving fully-compensated ferrimagnetism.  Very recently, fully-compensated ferrimagnetism has been experimentally realized in bilayer  $\mathrm{CrPS_4}$ by a perpendicular electric field, which  can switch the spin polarization of the conduction band on and off\cite{nn}.   This further confirms the feasibility of our proposal.
The electric field, in conjunction with the spin-orbital coupling (SOC), has the potential to enable MnSe to exhibit AVHE.  Additionally, we explore the induction of fully-compensated ferromagnetism through Se vacancies and Janus engineering.

 We conduct spin-polarized first-principles calculations within density functional theory (DFT)\cite{1}using the Vienna Ab Initio Simulation Package (VASP)\cite{pv1,pv2,pv3}.  The Perdew-Burke-Ernzerhof generalized gradient approximation (PBE-GGA)\cite{pbe}  is employed as the exchange-correlation functional. The calculations are performed with the kinetic energy cutoff  of 500 eV,  total energy  convergence criterion of  $10^{-8}$ eV, and  force convergence criterion of 0.001 $\mathrm{eV.{\AA}^{-1}}$. To account for the localized nature of Mn-3$d$ orbitals, a Hubbard correction $U_{eff}$=2.3 eV\cite{zg3}  is applied using the rotationally invariant approach proposed by Dudarev et al\cite{du}.
 A vacuum space of more than 15 $\mathrm{{\AA}}$ along the $z$-direction is introduced to prevent interactions between neighboring slabs. The BZ is sampled using a 15$\times$15$\times$1 Monkhorst-Pack $k$-point meshes  for crystal structure optimization and electronic structure calculations. The Berry curvatures are calculated directly from the wave functions using Fukui's method\cite{bm}, as implemented in the VASPBERRY code\cite{bm1,bm2,bm3}.

 \begin{figure}[t]
    \centering
    \includegraphics[width=0.45\textwidth]{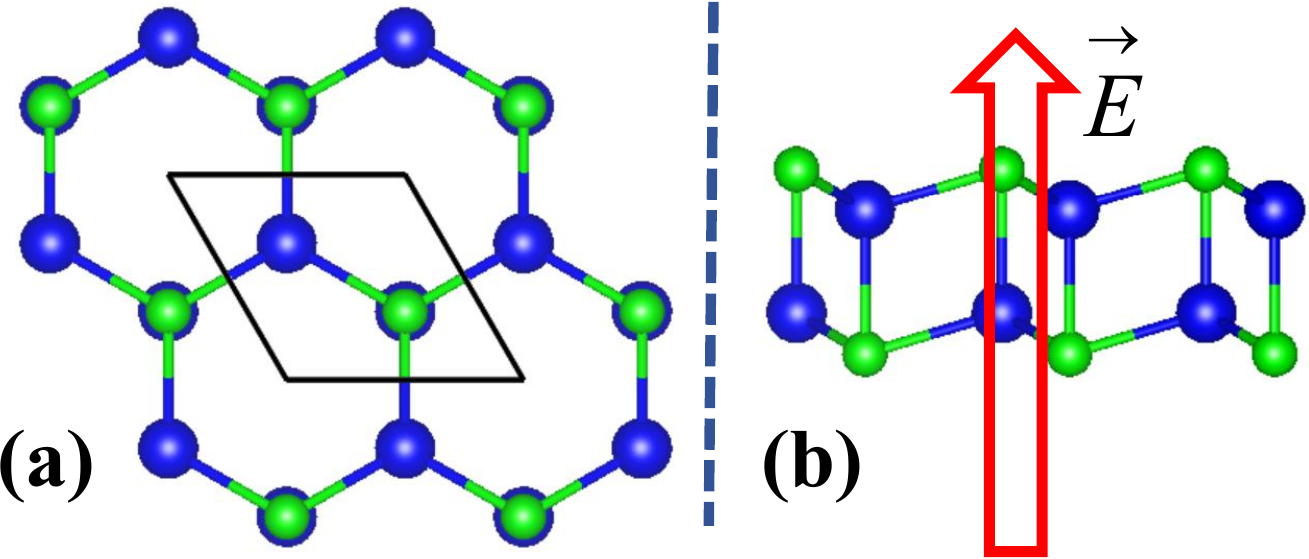}
    \caption{(Color online) For  monolayer  MnSe,  the top (a) and side (b) views of crystal structures. In (a, b),  the blue and green balls  represent Mn  and
      Se atoms, respectively. In (b), the red arrow represents the out-of-plane electric field.}\label{a}
\end{figure}

The monolayer MnSe has been synthesized in
experiment\cite{zg3}. As shown in \autoref{a}, the MnSe monolayer  includes
two buckled honeycomb MnSe sublayers, and they are interconnected by
Mn-Se bonds. MnSe monolayer  crystallizes in the  $P\bar{3}m1$ (No.~164),  with lattice $P$ symmetry.
It has been proved  that the upper and lower sublayers
are coupled antiferromagnetically through A-type AFM ordering\cite{zg3}. The optimized  equilibrium lattice constants are $a$=$b$=4.27 $\mathrm{{\AA}}$.
When considering magnetic ordering, MnSe monolayer possesses $PT$ symmetry as a $PT$-antiferromagnet, giving rise to spin-degenerate energy bands.

In order to achieve fully-compensated ferrimagnet, $PT$ symmetry of MnSe monolayer should be broken, and its total magnetic moment still remains zero.
When an out-of-plane electric field is applied, the lattice $P$ symmetry can be removed. In other words, an out-of-plane electric field can break the $PT$ symmetry by generating layer-dependent electrostatic potential, giving rise to fully-compensated ferrimagnetism.
  Both sublayers of MnSe possess built-in electric fields, which are oriented in opposite directions and cancel each other out, resulting in a zero-net electric field. Therefore, by perturbing the built-in electric field of one sublayer through Se vacancies, the equilibrium is disrupted, and a net built-in electric field can be generated, leading to fully-compensated ferrimagnetism.
In fact, the out-of-plane electric field can also be equivalently replaced by the built-in electric field achieved through Janus engineering, producing fully-compensated ferrimagnetism.

\begin{figure}[t]
    \centering
    \includegraphics[width=0.45\textwidth]{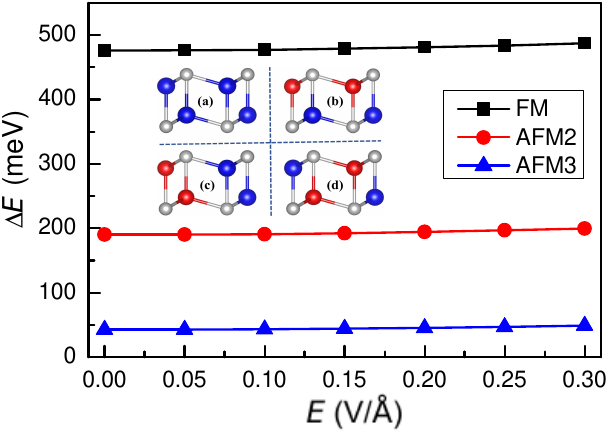}
    \caption{(Color online)For MnSe, the energy difference between  FM/AFM2/AFM3 and AFM1 orderings  as a function of $E$ (AFM1 ordering as a reference point). The inset shows four magnetic configurations, including FM (a), AFM1 (b), AFM2 (c) and AFM3 (d), and the blue, red and gray  balls  represent $\mathrm{Mn_{up}}$, $\mathrm{Mn_{dn}}$ and
      Se atoms, respectively. }\label{b}
\end{figure}

\begin{figure*}[t]
    \centering
    \includegraphics[width=0.96\textwidth]{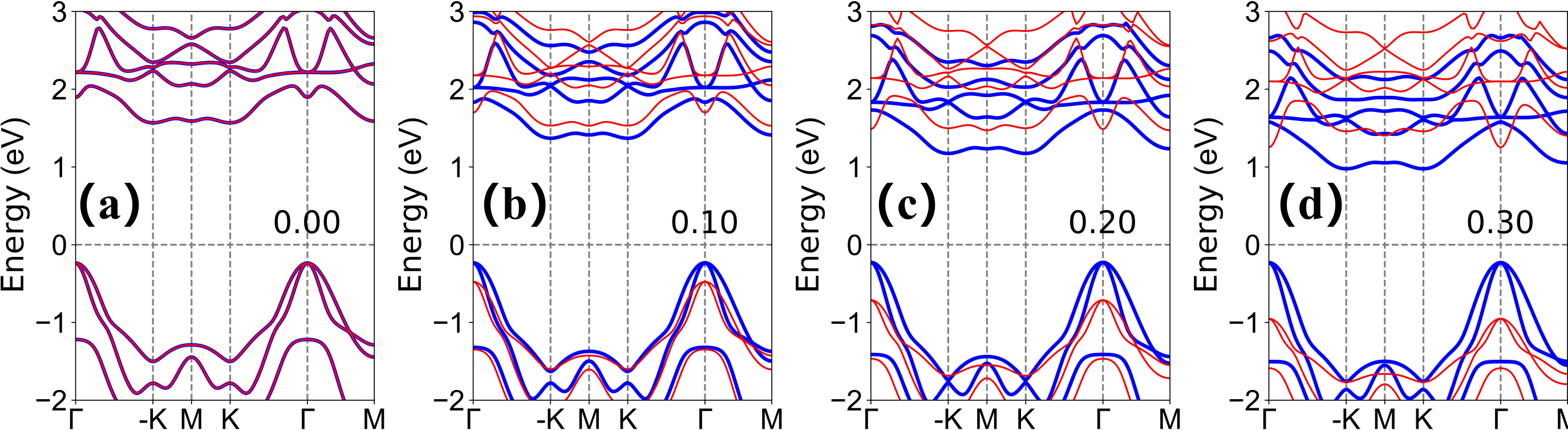}
     \caption{(Color online)For MnSe, the energy band structures without SOC at  representative $E$=+0.0 (a), +0.1 (b), +0.2 (c) and +0.3 (d)  $\mathrm{V/{\AA}}$.   In (a, b, c, d),  the blue (red) represents spin-up (spin-down) characters. }\label{c}
\end{figure*}
\begin{figure}[t]
    \centering
    \includegraphics[width=0.45\textwidth]{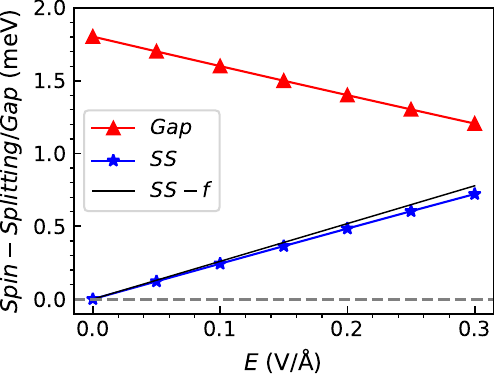}
     \caption{(Color online)For MnSe, the spin-splitting ($SS$) between first/second  and third/fourth valence bands at $\Gamma$ point and energy bandgap ($Gap$)  as a function of $E$. This black line  ($SS-f$) represents the linear fitting through the $eEd$.  }\label{d}
\end{figure}

Electric field can be an
 effective tool to break the lattice $P$ symmetry in 2D
 materials\cite{zg4}.  In order to achieve fully-compensated ferrimagnetism, magnetic ordering is another important factor. Therefore, we first determine the magnetic ground state of MnSe under the applied electric field.
As plotted in \autoref{b}, four magnetic configurations, including FM, AFM1, AFM2 and AFM3 orderings,  are considered, and AFM1 case of them is A-type AFM ordering.  The differences in
energy between  ferromagnetic (FM), AFM2, AFM3 and AFM1 states as a function of electric field $E$ are plotted  in \autoref{b}. Within considering $E$ range, MnSe monolayer always has AFM1 ordering, which provides the basic conditions to achieve fully-compensated ferrimagnetism.
With the increasing electric field, all energy differences increase, which means that the electric field can enhance magnetic interactions.

 The energy band structures of MnSe without SOC at  representative $E$=+0.0, +0.1, +0.2  and +0.3   $\mathrm{V/{\AA}}$ are shown in \autoref{c}.
\autoref{c} (a) shows the band structure without an electric field.
Due to $PT$ symmetry, the spin-up and spin-down bands are degenerate throughout the whole BZ.
It is  an indirect band gap semiconductor with gap value of 1.80 eV, which agrees well with available value of 1.76 eV\cite{zg5}.
The magnetic moments  (absolute value) of  Mn atoms of two sublayers are strictly equal due to lattice $P$ symmetry, and they are 4.38  $\mu_B$.

Upon the
application of an out-of-plane electric field, a layer-dependent electrostatic potential is produced to break $PT$ symmetry, and the spin degeneracy of
the bands is lifted. At  representative $E$=+0.3   $\mathrm{V/{\AA}}$,   the magnetic moments  of  two Mn atoms  are -4.35 $\mu_B$ and 4.38 $\mu_B$, and their absolute values not strictly equal due to broken $P$ symmetry. However, the total magnetic moment of MnSe  is  still equal to 0  $\mu_B$ in the presence of an electric field (The sum of  magnetic moments of Mn atom,  Se atoms and  interstitial region  is  zero.).
This compensation is strict, which is not  symmetry-driven, but rather due to the integer requirement of the magnetic moment caused by the energy gap\cite{f4}.
 Therefore, if the total magnetization of MnSe  is zero under an applied electric field,  it will  be strictly zero.  Electric field makes MnSe  to be a fully-compensated ferrimagnet. When the direction of  electric field is reversed, the order of spin splitting is also reversed (See FIG.S1\cite{bc}).

The spin-splitting  between first/second  and third/fourth valence bands at $\Gamma$ point and energy bandgap of MnSe  as a function of $E$ are plotted in \autoref{d}.  With increasing $E$,  the energy bandgap  decreases linearly, while spin-splitting increases linearly.
We also predict  the spin-splitting  by  $eEd$\cite{zg6}, where $e$  denotes the electron charge,  and the $d$  is the  difference in the average $z$-values of Mn and Se atoms of two MnSe sublayers (about 2.594 $\mathrm{{\AA}}$).  It is found that   the estimated spin-splitting is very close to the first-principle result. At  representative $E$=+0.3   $\mathrm{V/{\AA}}$, the spin-splitting can reach about 0.72 eV.  Recently, an intense
electric field larger than 0.4 $\mathrm{V/{\AA}}$  has been achieved  in 2D
materials by dual ionic gating\cite{zg7}, which provides the possibility of achieving a fully-compensated ferrimagnet with large spin-splitting in MnSe.

For  2D zero-net-magnetization magnets,  the fully-compensated ferrimagnet is a natural platform to realize the AVHE\cite{zg2}.
The orientation of magnetization plays a crucial role in generating spontaneous valley polarization. Specifically, only an out-of-plane magnetization direction can induce spontaneous valley splitting\cite{zg2}.  The magnetic orientation can be determined by calculating the magnetocrystalline anisotropy energy (MAE), which is given by the formula $E_{MAE}=E^{||}_{SOC}-E^{\perp}_{SOC}$. Here, $E^{||}_{SOC}$ and $E^{\perp}_{SOC}$  represent the energies when the spins are aligned in-plane and out-of-plane, respectively. The calculated MAE as a function of $E$ are plotted in FIG.S2\cite{bc}, and  the negative value indicates the in-plane magnetization of MnSe within considered $E$ range.

\begin{figure}[t]
    \centering
    \includegraphics[width=0.45\textwidth]{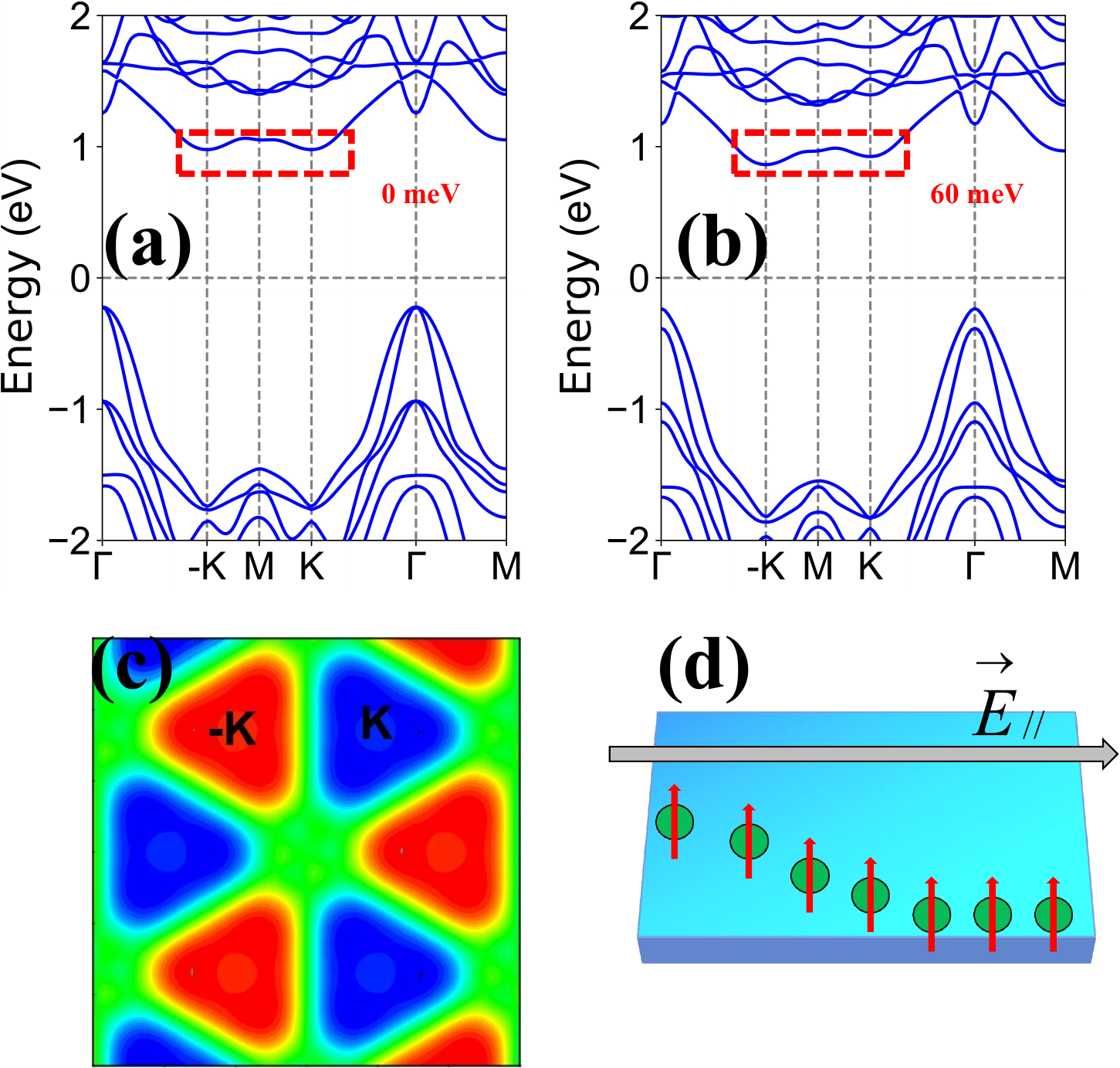}
     \caption{(Color online)For MnSe at $E$= +0.3 $\mathrm{V/{\AA}}$, (a and b): the energy band structures within SOC, including in-plane (a) and out-of-plane (b) magnetization directions; (c): for (b) case,  the distribution of Berry curvatures of   spin-up;  (d): for (b) case,  an appropriate  doping can produce AVHE   in the presence of a longitudinal in-plane electric field (gray arrow).  }\label{e}
\end{figure}

At $E$= +0.3 $\mathrm{V/{\AA}}$, the energy band structures with in-plane  magnetization are plotted \autoref{e} (a), which  shows no valley splitting.
To achieve valley polarization, the in-plane  magnetization should be tuned into  out-of-plane case.  In experiment, the manipulation of the $\mathrm{N\acute{e}el}$ vector can be effectively achieved through the application of spin-orbit torques (SOTs). Notably, experimental efforts have  realized the reorientation of $\mathrm{N\acute{e}el}$ vectors with switching angles of 90$^{\circ}$, 120$^{\circ}$ or 180$^{\circ}$\cite{zg8}.  At $E$= +0.3 $\mathrm{V/{\AA}}$, the energy band structures with out-of-plane  magnetization are shown in  \autoref{e} (b).   The   valley polarization with the valley splitting of 60 meV ($\Delta E_C=E_{K}^C-E_{-K}^C$) can be achieved in the conduction bands, and the energy of K valley
is higher than one of -K valley.

The -K and K valleys originate from the spin-up channel. The distribution of Berry curvatures for spin-up is depicted in \autoref{e} (c). It is evident that the Berry curvatures are opposite for the -K and K valleys. Upon applying a longitudinal in-plane electric field, the Bloch carriers acquire an anomalous transverse velocity $v_{\bot}$$\sim$$E_{\parallel}\times\Omega(k)$\cite{zg9}. When the Fermi level is positioned between the -K and K valleys of the conduction band, only the spin-up electrons  in the -K valley migrate to the boundary of the sample under the influence of an in-plane electric field, as shown in \autoref{e} (d), which  gives rise to AVHE\cite{zg10}.

\begin{figure}[t]
    \centering
    \includegraphics[width=0.45\textwidth]{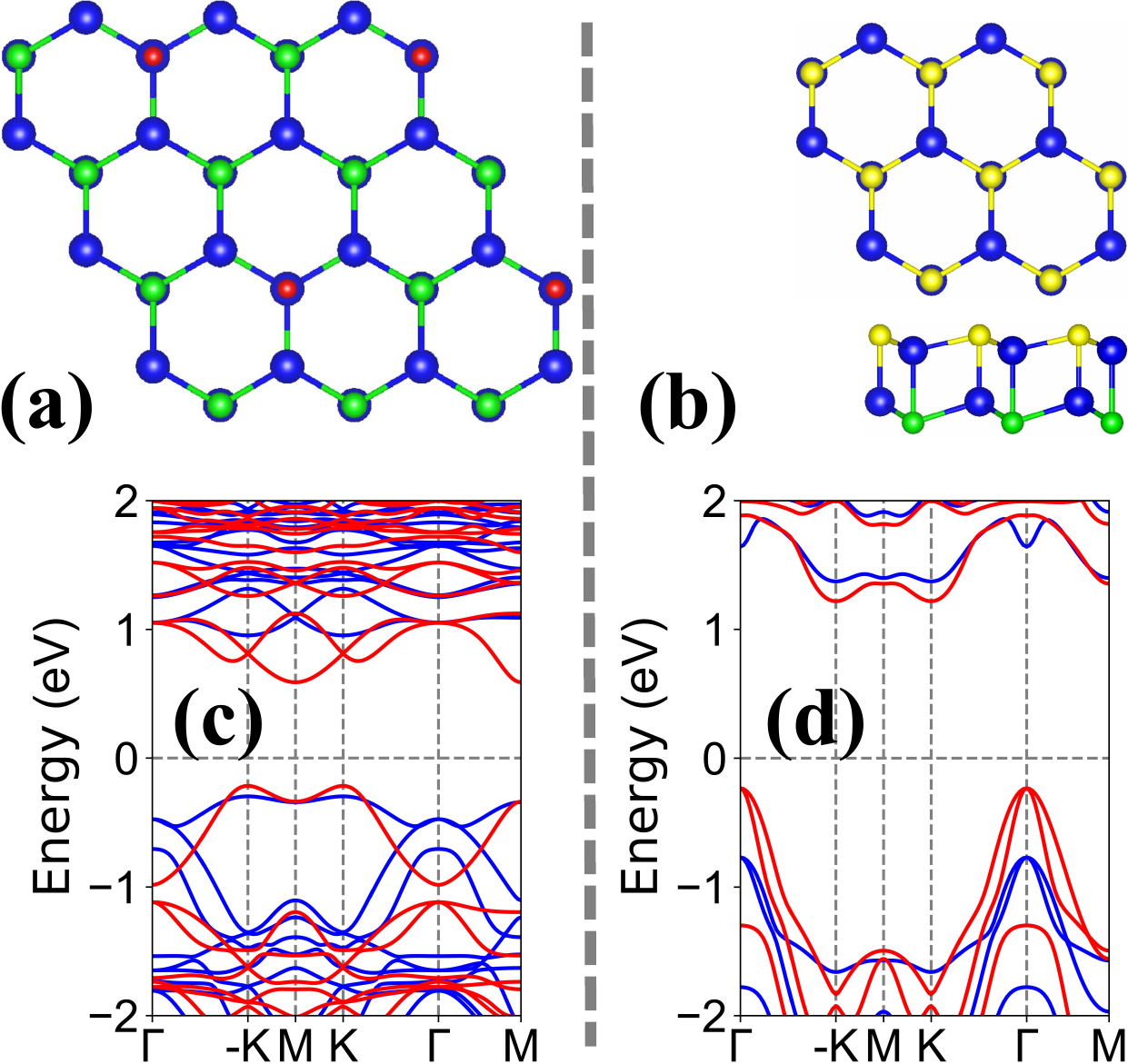}
     \caption{(Color online) (a and b):the  crystal structures of Se-vacancy MnSe and Janus  $\mathrm{Mn_2SSe}$ along with their energy band structures (c and d) without SOC. In (a, b),  the blue, green, yellow and red balls  represent Mn atoms, Se atoms, S atoms and Se vacancies, respectively.  In (c, d),  the blue (red) represents spin-up (spin-down) characters.} \label{f}
\end{figure}

In addition to the external electric field, vacancies may induce  fully-compensated ferrimagnetism in MnSe.
To simulate vacancies, we construct a 2$\times$2 supercell, in which one quarter of the Se atoms in upper sublayer are replaced by vacancies (see \autoref{f} (a)).
Calculated results show that the total magnetic moment of MnSe with Se-vacancy  is  still equal to 0  $\mu_B$, and its band structure shows significant spin-splitting (see \autoref{f} (c)).  These imply that vacancy  can indeed transform MnSe into a fully-compensated ferrimagnet.

In fact, the external electric field can be equivalently replaced by the built-in electric field, which can be achieved through Janus engineering.
The Janus   $\mathrm{Mn_2SSe}$ and   $\mathrm{Mn_2SeTe}$ have been predicted to possess A-type AFM ordering with good dynamic and thermal stabilities\cite{zg5}, and the crystal structures of $\mathrm{Mn_2SSe}$ are plotted in \autoref{f} (b).  Janus monolayer  crystallizes in the  $P3m1$ (No.~156),   and  the lack of horizontal mirror symmetry induces a built-in electric field.   The energy band structures of  Janus  $\mathrm{Mn_2SSe}$/$\mathrm{Mn_2SeTe}$ are plotted in \autoref{f} (d)/FIG.S3\cite{bc} ,  which is similar to that of MnSe with  external electric field, showing obvious spin-splitting.  The net magnetic moment of Janus monolayers still remains zero.  These show that  Janus   $\mathrm{Mn_2SSe}$ and   $\mathrm{Mn_2SeTe}$ are fully-compensated ferrimagnets.
When considering the SOC, Janus monolayers  with out-of-plane magnetization can simultaneously exhibit valley-splitting (FIG.S4\cite{bc})  and spin-splitting (\autoref{f} (d)), which can lead to the realization of AVHE.

In summary, we  explore the potential of achieving fully-compensated ferrimagnetism based on the A-type $PT$-antiferromagnet MnSe as a parent material. By applying an out-of-plane electric field,  the  lattice $P$ symmetry is removed, thereby breaking the $PT$ symmetry and inducing spin-splitting. This manipulation, combined with SOC, enables MnSe to exhibit AVHE. Our findings provide experimentally feasible approaches to realizing fully-compensated ferrimagnets, thereby advancing the development of 2D magnetic materials with enhanced performance for miniaturization, ultradensity, and ultrafast applications.

~~~~\\
\textbf{Conflicts of interest}
\\
There are no conflicts to declare.

~~~~~\\

\begin{acknowledgments}
This work is supported by Natural Science Basis Research Plan in Shaanxi Province of China  (2021JM-456). We are grateful to Shanxi Supercomputing Center of China, and the calculations were performed on TianHe-2.
\end{acknowledgments}

\end{document}